\documentclass[prd,twocolumn,nofootinbib]{revtex4}
\usepackage{amsmath,amssymb,epsf,graphicx}

\def\lsim{\mathrel{\hbox{\rlap{\hbox{\lower4pt\hbox{$\sim$}}}\hbox{$<$}}}}
\def\gsim{\mathrel{\hbox{\rlap{\hbox{\lower4pt\hbox{$\sim$}}}\hbox{$>$}}}}
\def\be{\begin{equation}}
\def\ee{\end{equation}}
\def\bea{\begin{eqnarray}}
\def\eea{\end{eqnarray}}

\begin{document}


\title{Constraints on Cosmic Super-Strings from Kaluza-Klein Emission}
\date{\today}
\author{Jean-Fran\c{c}ois Dufaux}
\affiliation{APC, Univ. Paris Diderot, CNRS/IN2P3, CEA/Irfu, Obs. de Paris, Sorbonne Paris Cit\'e, France}

\begin{abstract}
Cosmic super-strings interact generically with a tower of light and/or strongly coupled Kaluza-Klein (KK) modes associated with the geometry of the internal space. We study the production of KK particles by cosmic super-string loops, and show that it is constrained by Big Bang Nucleosynthesis (BBN). We study the resulting constraints in the parameter space of the underlying string theory model and highlight their complementarity with the regions that can be probed by current and upcoming gravitational wave (GW) experiments.
\end{abstract}

\maketitle

\noindent

Cosmic strings are linear concentrations of energy that appear in several high-energy physics scenarios and may have a variety of cosmological and astrophysical consequences~\cite{VS}. Field theory cosmic strings have been abundantly studied since the 70's, but the subject has been recently regenerated by the realization that cosmic strings occur also as fundamental objects of string theory, see \cite{pol} for reviews. This opens the opportunity that these cosmic super-strings could lead to observational signatures of string theory, if they can be distinguished to some degree from their field-theory counterparts. In this letter, we study one specific aspect of cosmic super-strings that may help this distinction: their production of KK particles associated with the extra dimensions. 

The extra dimensions play a crucial role for cosmic super-strings, being responsible for making their tension $\mu$
small enough compared to the Planck scale, as required by observations~\cite{pol}. This can be achieved in two different ways. The first one is to have cosmic strings localized at the bottom of a warped internal space, called throat~\cite{KS}. 
Relatively few KK modes are then lighter than $\sqrt{\mu}$, but they are strongly coupled to the cosmic strings because their wave-function is also localized at the bottom of the throat~\cite{RS}. The other possibility is to have a flat internal space with large volume~\cite{ADD}. The KK modes are then only gravitationally coupled to the cosmic strings, but many of them are very light. Whether KK modes are strongly coupled or very light, they can be efficiently produced by cosmic strings in both cases. Here we will focus on the first case (warped throat), where the most explicit models have been developed. The other case leads to similar results~\cite{prep}.

The tension of cosmic super-strings is highly model-dependent. When they are produced at the end of brane inflation, 
and in models where the Cosmic Microwave Background anisotropies are generated only by the quantum fluctuations of a 
single inflaton field, their tension is expected to be in the approximate range $10^{-13} \lsim G \mu \lsim 10^{-7}$ ($G$ is Newton's constant)~\cite{pol}. The gravitational effects of such cosmic strings can be observable in the near future, in particular with GW experiments~\cite{DV, DVsuper, siemens, elisa}. In general, however, cosmic super-strings may have a much smaller tension. This occurs in particular when they are produced in models of brane inflation at lower energy scales, see e.g.~\cite{lev} where $G \mu$ can be as small as $10^{-34}$. This also occurs when they form at Hagedorn phase transitions~\cite{MD} after inflation, e.g. in multi-throat scenarios~\cite{pol,chen}. Because the gravitational effects of such light cosmic strings are much weaker, it is crucial to look for their other possible signatures, in particular the production of (massive) particles. Stability constraints significantly restrict the possible interactions of cosmic super-strings~\cite{pol}. However, they always couple to the ten-dimensional metric and the corresponding KK modes. We will show that this leads to constraints on cosmic super-strings in ranges of string tensions that are complementary to the ones that can be probed with GW experiments.

We assume for simplicity that only F-strings are present or dominate the loop number density. The generalization to 
D-strings is straightforward~\cite{prep}. When different kinds of strings are present and interact, we expect the F-strings to dominate the loop number density, because they typically have the smallest reconnection probabilities~\cite{psuper}. The Nambu-Goto action for a cosmic F-string in a ten-dimensional space-time with metric $\tilde{g}_{A B}$ ($A, B = 0, ..., 9$) reads
\be
\label{action}
\frac{1}{2\pi\,\alpha'}\,\int d\tau d\sigma\,\sqrt{-\tilde{\gamma}} \, ,
\ee
where $1 / (2 \pi\,\alpha')$ is the fundamental string tension and $\tilde{\gamma}$ is the determinant of the 
two-dimensional metric induced on the string worldsheet: 
$\tilde{\gamma}_{\alpha \beta} = \tilde{g}_{AB}\,\partial_{\alpha} X^A\,\partial_{\beta} X^B$ with $\alpha, \beta = 0, 1$ 
and the embedding of the worldsheet in the ten-dimensional spacetime is parametrized as $x^A = X^A(\tau, \sigma)$. The 
ten-dimensional metric can be written as 
$\tilde{g}_{AB} dx^A dx^B = e^{2 A (\bar{y})}\,\eta_{\mu \nu} dx^{\mu} dx^{\nu} + ...$ where $\eta_{\mu \nu}$ is the 
four-dimensional Minkowski metric ($\mu, \nu = 0, ..., 3$), $x^{\mu}$ are coordinates in the four non-compact dimensions and $\bar{y}$ denotes coordinates in the internal space. As emphasized in \cite{psuper}, the position of the cosmic strings in the internal space corresponds to worldsheet moduli that are not protected by any symmetry and should be fixed at the minimum of their effective potential: $X^A(\tau, \sigma) = \mathrm{constant}$ at the classical level for $A = 4, ..., 9$. The action (\ref{action}) then reduces to the usual four-dimensional Nambu-Goto action with the induced metric 
$\gamma_{\alpha \beta} = \eta_{\mu \nu}\,\partial_\alpha X^\mu\,\partial_\beta X^\nu$ and the effective tension 
\be
\label{muwarp}
\mu = \frac{e^{2 A_b}}{2 \pi\,\alpha'} \equiv \frac{M_s^2}{2 \pi} \, ,
\ee
where $e^{A_b} \ll 1$ is the warp factor at the bottom of the throat (where the cosmic super-strings are located) and 
$M_s = e^{A_b} / \sqrt{\alpha'}$ is the local string mass scale there.

We will focus on KK modes that correspond to spin-$2$ fields in four dimensions, because these are the simplest and most generic ones. These modes are always present and the massless one corresponds to the usual four-dimensional graviton. Their coupling to the cosmic string can be obtained from the previous equations with the replacement 
$\eta_{\mu \nu} \rightarrow \eta_{\mu \nu} + \sum_{\bar{n}} \Phi_{\bar{n}}(\bar{y})\,h^{\bar{n}}_{\mu \nu}(x)$, where 
$\bar{n}$ denotes collectively the mode numbers associated to each KK modes, $\Phi_{\bar{n}}(\bar{y})$ is their wavefunction in the internal space (normalized in order to obtain a canonical kinetic term for $h^{\bar{n}}_{\mu \nu}$ after dimensional reduction), and $\partial^{\mu} h^{\bar{n}}_{\mu \nu} = \eta^{\mu \nu} h^{\bar{n}}_{\mu \nu} = 0$. The term that is linear in $h^{\bar{n}}_{\mu \nu}$ in the action (\ref{action}) can then be written as
\be
\label{int2}
\sum_{\bar{n}} \frac{\lambda_{\bar{n}}}{\sqrt{\mu}}\,\int d^4x\,h^{\bar{n}}_{\mu \nu}\,T^{\mu \nu} \, ,
\ee
where $T_{\mu \nu}$ is the four-dimensional energy-momentum tensor of the string~\cite{VS} (located at 
$\bar{y} = \bar{y}_b$) and we have defined $\lambda_{\bar{n}} = \sqrt{\mu}\,\Phi_{\bar{n}}(\bar{y}_b) / 2$. 

The properties of the spin-$2$ KK modes in a warped throat are well understood, see e.g.~\cite{KKrelics} and references therein. Their masses are quantized as
\be
\label{mwarp}
m_{\bar{n}} = \xi_{\bar{n}}\,\frac{e^{A_b}}{R} \, , 
\ee
where the $\xi_{\bar{n}}$ are numerical coefficients and $R > \sqrt{\alpha'}$ is the radius of curvature at the bottom of the throat. The lowest-lying KK modes have $\xi_{\bar{n}} \sim 1$ and are lighter than the string scale, 
$m_{\bar{n}} < M_s$. The massive modes are strongly localized at the bottom of the throat, where the amplitude of their normalized wave-function $\Phi_{\bar{n}}(y_b)$ can be found in \cite{KKrelics}. This gives 
\be
\label{phim}
\lambda_{\bar{n}} \simeq (2 \pi)^3\,g_s\,\left(\frac{\sqrt{\alpha'}}{R}\right)^3 \, ,
\ee
where $g_s$ is the fundamental string coupling and we have used Eq.~(\ref{muwarp}). On the other hand, the massless mode 
$\bar{n} = \bar{0}$ has a constant wave-function in the internal space and $\lambda_{\bar{0}} = \sqrt{8 \pi G\,\mu}$. 

From the interaction (\ref{int2}), the energy emitted in massive KK modes with mode numbers $\bar{n}$ by a cosmic super-string loop can be calculated as
\be
\label{Em}
E_{\bar{n}} = \frac{\lambda_{\bar{n}}^2}{2 \mu} \int \frac{d^3\mathbf{k}}{(2 \pi)^3}
\left(T^{\mu\nu}(\omega_k, \mathbf{k}) T_{\mu\nu}^{*}(\omega_k, \mathbf{k}) - 
\frac{1}{3} |T^{\lambda}_{\lambda}(\omega_k, \mathbf{k})|^2\right) 
\ee
where $T_{\mu\nu}(\omega_k, \mathbf{k}) = \int d^4x\,T_{\mu\nu}(t, \mathbf{x})\,e^{i k_\lambda x^\lambda}$ is the (double) Fourier transform of the loop energy-momentum tensor and we have introduced the $4$-vector 
$k^\lambda = (\omega_k, \mathbf{k})$ with $\omega_k = \sqrt{k^2 + m_{\bar{n}}^2}$ and $k = |\mathbf{k}|$. The massive 
spin-$2$ modes have five physical degrees of freedom while the massless one has only two. For the latter, the $1/3$ factor in (\ref{Em}) must be replaced by $1/2$ and we recover the standard expression for the GW energy emitted by a source. In the following, we will be interested in the massive modes. 

The production of massive scalar particles by cosmic string loops has been studied by several authors~\cite{scalar,moduli}. The calculation in our case is similar, except that we deal with massive spin-$2$ fields and that we have a tower of them. For loops of invariant length $L \gg 1 / m_{\bar{n}}$, we calculate $T_{\mu\nu}(\omega_k, \mathbf{k})$ in the stationary phase approximation as in~\cite{DV}. The dominant contribution comes from cusps, which emit massive particles with a typical energy $\omega_k \simeq k \sim m_{\bar{n}}\,\sqrt{m_{\bar{n}} L} \gg m_{\bar{n}}$ inside a cone of opening angle 
$\theta \sim 1 / \sqrt{m_{\bar{n}} L}$ in the rest frame of the loop~\cite{scalar}. The full $\mathbf{k}$-dependence of 
$T_{\mu\nu}(\omega_k, \mathbf{k})$ can be calculated analytically and the integral in (\ref{Em}) can then be performed numerically~\cite{prep}. This gives
\be
\label{En}
E_{\bar{n}} \, \approx \, \frac{\lambda_{\bar{n}}^2}{2} \, \mu \, \sqrt{\frac{L}{m_{\bar{n}}}} \, .
\ee

Neglecting the production of fundamental string states, we must require $m_{\bar{n}} < M_s$ in the above calculation in order to trust the Nambu-Goto effective description. The total energy emitted by a cusp in massive KK modes can then be calculated as
\be
\label{Etot2}
E \, = \, \sum_{\bar{n}}^{m_{\bar{n}} < M_s} E_{\bar{n}} \, \approx \, \frac{\kappa}{2}\,g_s^{2}\,\mu^{3/4}\,\sqrt{L} \, ,
\ee
where we have used Eqs.~(\ref{muwarp}, \ref{mwarp}, \ref{phim}) and we define
\be
\label{kappaE}
\kappa \equiv (2 \pi)^{23/4}\,\left(\frac{\sqrt{\alpha'}}{R}\right)^{11/2}\,
\sum_{\bar{n}}^{m_{\bar{n}} < M_s} \xi_{\bar{n}}^{-1/2} \, .
\ee
The coefficient $\kappa$ depends on the details of the compactification and of the KK spectrum. In the throat model that 
we will consider below, we will find $\kappa \approx 10$ for a range of values of the ratio 
$R / \sqrt{\alpha'}$. When this ratio increases, the factor in front of the the sum in (\ref{kappaE}) decreases, but more terms must be included in the sum. These two effects tend to compensate each other.  

The power emitted in massive KK modes by a cosmic super-string loop of length $L$ can be calculated as $P_{KK} = c\,E / T$, where $c$ is the average number of cusps on loops per oscillation period (expected to be $c \sim 1$) and $T = L / 2$ is the period of oscillation. Taking $c = 1$, 
\be
\label{Pkk}
P_{KK} = \Gamma_{KK}\,\frac{\mu^{3/4}}{\sqrt{L}} \hspace*{0.3cm} \mbox{with} \hspace*{0.3cm} 
\Gamma_{KK} \approx \kappa\,g_s^{2} \, .
\ee
Interestingly, this result is comparable to the power emitted in the constituent fields of standard Abelian-Higgs cosmic strings in the process of cusp annihilation~\cite{BPO}. Note however that we obtained (\ref{Pkk}) within the regime of validity of the Nambu-Goto description, while cusp annihilation occurs beyond this regime. We will also see that KK emission by cosmic super-strings may have specific cosmological consequences. 

Before doing so, we must consider the effect of KK emission on the loop number density. Once the loops are produced, they shrink by emitting both GW, as usual, but also KK modes - the respective powers are given by $P_{GW} = \Gamma G \mu^2$ with $\Gamma \sim 50$ and by $P_{KK}$ in (\ref{Pkk}). KK emission then decreases the lifetime of the loops and thus also their number density. A detailed study of this effect will appear elsewhere. Here we will only give the results that we will need in the following. We will focus on the case where the loops are produced from the long string network with a large initial size, 
$L = \alpha\,t$ with $\alpha \approx 0.1$, as indicated by the most recent simulations - see \cite{simul} and references therein. In the radiation era, the number density of loops with lengths between $L$ and $L + dL$ at time $t$ can then be written as
\be
\label{nL}
n_L(t) dL \approx \frac{\zeta \sqrt{\alpha} dL}{p L^{5/2} 
t^{3/2}} \hspace*{0.3cm} \mbox{for} \hspace*{0.3cm} L_H(t) < L < \alpha t \, ,
\ee
where $\zeta \approx 10$ and $L_H(t) = \Gamma G \mu t$ or $(\Gamma_{KK} t)^{2/3} / \mu^{1/6}$, whichever is larger. The loops with $L = L_H(t)$ have a lifetime of the order of the Hubble time and are the most abundant ones. For cosmic 
super-strings, the reconnection probability $p$ can be smaller than unity. This is usually expected to increase the loop number density as $n_L \propto p^{-1}$~\cite{DVsuper} (see however \cite{avgou}) and we took this factor into account in Eq.~(\ref{nL}). Except for this factor, Eq.~(\ref{nL}) reduces to the standard result for loops emitting only GW~\cite{VS} when $t > t_{\mu} \equiv \Gamma^2_{KK} \sqrt{\Gamma G} / (\Gamma G \mu)^{7/2}$. For times and string tensions such that 
$t < t_{\mu}$, most of the loops decay by emitting mainly KK modes and their number density is reduced.

Once produced by cosmic super-strings, the KK modes should decay relatively quickly into Standard Model (SM) fields. Indeed, because the KK modes are light, they are also abundantly produced in the early universe at high temperatures. For instance, they are the main decay products of brane annihilation at the end of brane inflation and must then efficiently decay into SM fields in order to reheat the universe, see~\cite{KKrelics} and references therein. These KK modes produced at early times must decay sufficiently early to avoid cosmological problems~\cite{taur}. On the other hand, cosmic super-strings continue to emit KK modes at later times and their decay is then severely constrained. The strongest~\cite{BBNTD,prep} constraint in our case comes form the photo-dissociation of BBN light 
elements~\cite{BBN, diffuse}, which depends on the total energy density injected in the cosmological medium in one Hubble time. This can be estimated as $\Delta \rho_{inj} \approx t\,\dot{\rho}_{KK}$ where 
$\dot{\rho}_{KK} = \int dL \, n_L(t) \, P_{KK}$~\cite{instdec}. Using Eqs.~(\ref{Pkk},\ref{nL}), this gives
\be
\frac{\Delta \rho_{inj}}{s} \approx \frac{10 \Gamma_{KK} \zeta \sqrt{\alpha}}{p \Gamma^2 (G \mu)^{5/4} t} 
\, \mathrm{Min}\left[1 , \frac{t^{2/3}}{t^{2/3}_{\mu}}\right] 
\, ,
\ee
where $s \sim 0.1 / (t^{3/2} G^{3/4})$ is the entropy density at the time of interest and 
$\mathrm{Min}[a , b] = a$ or $b$, whichever is smaller. Photo-dissociation constraints imply 
$\Delta \rho_{inj} / s \lsim 10^{-14}$ GeV at $t \approx 10^8$ sec, see Fig.~42 of \cite{BBN}.

\begin{figure}[hbt]
\centering
\includegraphics[width=\columnwidth]{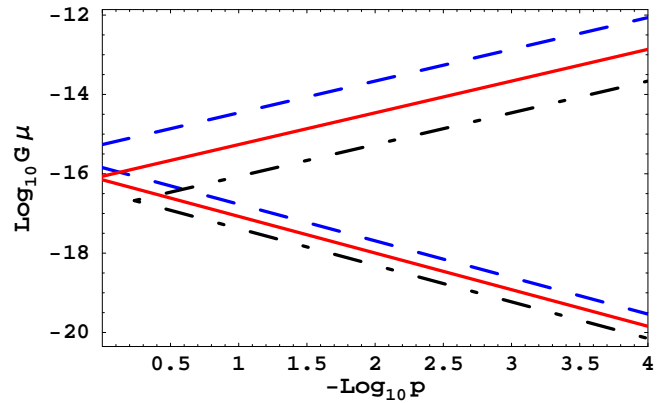}
\vspace*{-0.4 cm}
\caption{BBN (photo-dissociation of light elements) constraints from KK emission for $\zeta = 10$, $\alpha = 0.1$, 
$\Gamma = 50$, and $\Gamma_{KK} = 10$ (dashed lines), $1$ (plain lines) and $0.1$ (dot-dashed lines). In each case, the region between the two corresponding lines is excluded.}
\label{pGmu}
\end{figure}

The resulting constraints on cosmic super-strings in the ($p , G \mu$)-plane are shown in Fig.~\ref{pGmu}. For $p = 1$ 
and $\Gamma_{KK} \sim 1$, only a very small range of string tensions might be constrained. When $p$ decreases, 
the constraints become quickly stronger. For a given $p$, only a range of tensions is constrained. In the upper half of this range, the loops decay mainly by GW emission. Increasing $G \mu$ then decreases the loop lifetimes and their number density, so sufficiently large tensions are not constrained. On the other hand, in the lower half of the range, the loops decay mainly by KK emission. In this regime, the loop number density \emph{decreases} when $G \mu$ decreases, so sufficiently small tensions are not constrained either. 

\begin{figure}[hbt]
\centering
\includegraphics[width=\columnwidth]{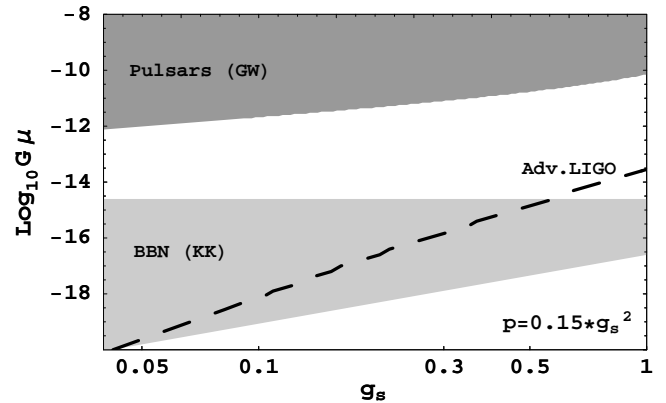}
\vspace*{-0.4 cm}
\caption{BBN constraints from KK emission compared to the parameter space that can be probed by GW experiments 
(adapted from \cite{elisa}) for F-strings in a KS throat with $\zeta = 10$, $\alpha = 0.1$, $\Gamma = 50$ and 
$R / \sqrt{\alpha'} \sim 10$ (see the main text for details). The upper shaded region is excluded by current pulsar data (GW), the lower one by BBN (KK), and the region above the dashed line is accessible to Advanced LIGO (GW).} 
\label{gsGmu}
\end{figure}

In a given model, the parameters $\Gamma_{KK}$ and $p$ are in general not independent, because they both depend on $g_s$ 
and on the compactification details, in particular on $R / \sqrt{\alpha'}$. To illustrate this, we now consider the most studied example of throat geometry, which is based on the Klebanov-Strassler (KS) solution~\cite{KS}. The reconnection probabilities in that case were studied in \cite{psuper}. The results of that paper show that, except for 
$1 < R / \sqrt{\alpha'} \lsim 4$, $p$ depends only logarithmically on $R / \sqrt{\alpha'}$ for F-strings, with 
$p \approx 0.15 g_s^2$ for $R / \sqrt{\alpha'} \sim 10$. To estimate $\kappa$ in (\ref{kappaE}), we use Table 1 of 
\cite{KKrelics} for the spectrum of the lowest-lying spin-$2$ KK modes in a KS throat, approximated as a patch of 
$AdS_5 \times T^{1,1}$. We find that $\kappa$ depends also very weakly on $R / \sqrt{\alpha'}$, with $\kappa \approx 10$ for $R / \sqrt{\alpha'} \sim 10$. Eq.~(\ref{Pkk}) then gives $\Gamma_{KK} \approx 10 g_s^2$. The BBN constraints on the string tension for these values of $\Gamma_{KK}$ and $p$ are shown in Fig.~\ref{gsGmu} as a function of $g_s$. We also show in Fig.~\ref{gsGmu} the regions of the parameter space for which the stochastic GW background from the strings is accessible to current pulsar timing observations and to Advanced LIGO, as discussed in \cite{elisa}.

We see from Fig.~\ref{gsGmu} that the combination of the current constraints coming from GW (pulsars) and KK (BBN) emissions leads already to stringent limits on $G \mu$. For instance, for $g_s \sim 0.05$, $G \mu$ is restricted to the ranges $10^{-15} \lsim G \mu \lsim 10^{-12}$ or $G \mu \lsim 10^{-20}$. Furthermore, the whole region between the two
shaded areas in Fig.~\ref{gsGmu}, which is currently unconstrained, will be accessible to either Advanced LIGO or the ESA space-interferometer eLISA~\cite{elisa}. If no cosmological GW background is found by these experiments, the lower value of $G \mu$ that is constrained by KK emission would then set the strongest upper bound on the string tension, which reads 
$G \mu \lsim 2.5 \times 10^{-17} \, g_s^{32/13}$ for the model shown in Fig.~\ref{gsGmu}. The best outcome, however, is of course that these experiments do observe a stochastic GW background. Unfortunately, the predictions for the GW spectrum from cosmic super-strings are highly degenerate in the $(p, G \mu)$-plane. Therefore, even in the most optimistic case where we can gain confidence in the fact that the observed background is produced by cosmic strings, it will not be possible to determine the fundamental parameters. The constraints from KK emission will then allow to discriminate between a wide range of possibilities, see Fig.~\ref{gsGmu}. 

On the string theory side, these constraints are sensitive to $g_s$ and to the geometry of the internal space. Although we focused on the most developed KS throat model, the calculation can be generalized to other string theory models. The constraints depend also on the typical size of the loops when they are produced, on the way $n_L$ varies with $p$, and on the average number of cusps per loop oscillation. These properties are still uncertain and can hopefully be determined by the next generation of simulations. We expect however our main conclusion - the complementarity between KK and GW emissions in probing the parameter space of cosmic super-strings - to be qualitatively robust. This may ultimately allow to constrain or determine fundamental parameters of string theory, such as $g_s$ and the compactification details. Meanwhile, other possible consequences of KK emission remain to be studied, e.g. the production of cosmic rays or dangerous relics.




\begin{thebibliography}{10}

\bibitem{VS}
  A.~Vilenkin and E.~P.~S.~Shellard, \emph{Cosmic Strings and Other Topological Defects} 
  (Cambridge University Press, Cambridge, 1994).

\bibitem{pol} 
  J.~Polchinski,
  arXiv:hep-th/0412244.
  E.~J.~Copeland, L.~Pogosian and T.~Vachaspati,
  Class.\ Quant.\ Grav.\  {\bf 28}, 204009 (2011).

\bibitem{KS}
  I.~R.~Klebanov and M.~J.~Strassler,
  JHEP {\bf 0008}, 052 (2000).
S.~B.~Giddings, S.~Kachru and J.~Polchinski,
  Phys.\ Rev.\ D {\bf 66}, 106006 (2002).

\bibitem{RS}
  L.~Randall and R.~Sundrum,
  Phys.\ Rev.\ Lett.\  {\bf 83}, 3370 (1999).

\bibitem{ADD}
  N.~Arkani-Hamed, S.~Dimopoulos and G.~R.~Dvali,
  Phys.\ Lett.\  B {\bf 429}, 263 (1998).
V.~Balasubramanian, P.~Berglund, J.~P.~Conlon and F.~Quevedo,
  JHEP {\bf 0503}, 007 (2005).

\bibitem{prep}
  J.~F.~Dufaux,
  arXiv:1201.4850 [hep-th].

\bibitem{DV}
  T.~Damour and A.~Vilenkin,
  Phys.\ Rev.\ Lett.\  {\bf 85}, 3761 (2000) ; 
  Phys.\ Rev.\  D {\bf 64}, 064008 (2001).  

\bibitem{DVsuper}
  T.~Damour and A.~Vilenkin,
  Phys.\ Rev.\  D {\bf 71}, 063510 (2005).

\bibitem{siemens}
  X.~Siemens, V.~Mandic and J.~Creighton,
  Phys.\ Rev.\ Lett.\  {\bf 98}, 111101 (2007). 

\bibitem{elisa}
P.~Binetruy, A.~Bohe, C.~Caprini and J.~-F.~Dufaux,
  JCAP {\bf 1206}, 027 (2012).

\bibitem{lev}
  L.~Kofman and S.~Mukohyama,
  Phys.\ Rev.\  D {\bf 77}, 043519 (2008).

\bibitem{MD}
  M.~Majumdar and A.~Christine-Davis,
  JHEP {\bf 0203}, 056 (2002).

\bibitem{chen}
  X.~Chen,
  JHEP {\bf 0508}, 045 (2005).

\bibitem{psuper}
  M.~G.~Jackson, N.~T.~Jones and J.~Polchinski,
  JHEP {\bf 0510}, 013 (2005).

\bibitem{KKrelics}
  J.~F.~Dufaux, L.~Kofman and M.~Peloso,
  Phys.\ Rev.\  D {\bf 78}, 023520 (2008).

\bibitem{scalar}
  M.~Srednicki and S.~Theisen,
  Phys.\ Lett.\  B {\bf 189}, 397 (1987).
  T.~Vachaspati,
  Phys.\ Rev.\  D {\bf 81}, 043531 (2010).
  V.~Berezinsky, E.~Sabancilar and A.~Vilenkin,
  Phys.\ Rev.\  D {\bf 84}, 085006 (2011).

\bibitem{moduli}
T.~Damour and A.~Vilenkin,
  Phys.\ Rev.\ Lett.\  {\bf 78}, 2288 (1997).
  M.~Peloso and L.~Sorbo,
  Nucl.\ Phys.\  B {\bf 649}, 88 (2003).
E.~Babichev and M.~Kachelriess,
  Phys.\ Lett.\ B {\bf 614}, 1 (2005).
E.~Sabancilar,
  Phys.\ Rev.\ D {\bf 81}, 123502 (2010).

\bibitem{BPO}
  J.~J.~Blanco-Pillado and K.~D.~Olum,
  Phys.\ Rev.\  D {\bf 59}, 063508 (1999). 
  K.~D.~Olum and J.~J.~Blanco-Pillado,
  Phys.\ Rev.\  D {\bf 60}, 023503 (1999).

\bibitem{simul}
  J.~J.~Blanco-Pillado, K.~D.~Olum and B.~Shlaer,
  Phys.\ Rev.\  D {\bf 83}, 083514 (2011).

\bibitem{avgou}
  A.~Avgoustidis and E.~P.~S.~Shellard,
  Phys.\ Rev.\  D {\bf 73}, 041301 (2006).

\bibitem{taur}
  This requires the lifetime (at rest) $\tau$ of the KK modes to be certainly smaller than $1$ sec from BBN, and often much smaller for baryogenesis.

\bibitem{BBNTD}
  G.~Sigl, K.~Jedamzik, D.~N.~Schramm and V.~S.~Berezinsky,
  Phys.\ Rev.\  D {\bf 52}, 6682 (1995).

\bibitem{BBN}
  M.~Kawasaki, K.~Kohri and T.~Moroi,
  Phys.\ Rev.\  D {\bf 71}, 083502 (2005).

\bibitem{diffuse}
Observations of the diffuse gamma-ray background lead to similar constraints and the results are not affected 
by a possible friction-dominated epoch~\cite{prep}.

\bibitem{instdec}
This assumes that the KK modes decay in less than one Hubble time despite their large initial boost, $\gamma \tau \lsim t$ where $\tau$ is their lifetime at rest and $\gamma \sim \sqrt{m_{\bar{n}} L} \sim \mu^{1/4} L^{1/2}_H(t)$. This is a good approximation for the range of string tensions and at the epoch ($t \sim 10^8$ sec) that we will consider except if $\tau$ is very large, close to the upper bound $\tau \sim 1$ sec~\cite{taur}. In that case, the KK modes decaying at $t \sim 10^8$ sec would be produced at $t_p \ll t$ and $\Delta \rho_{inj}(t) \sim t_p \, \dot{\rho}_{KK}(t_p) \, a^4(t_p) / a^4(t)$ instead of $t\,\dot{\rho}_{KK}(t)$. This would \emph{tighten} the constraints we will obtain. 

\end{thebibliography}
\end{document}